\documentclass[intlimits,twoside,a4paper]{article}
\usepackage{amsmath,amssymb}
\usepackage{graphicx}
\usepackage{wrapfig}
\usepackage[T2A]{fontenc}
\usepackage[cp1251]{inputenc}

\usepackage{cmpj3}

\issue{2018}{21}{4}{43001}
\doinumber{10.5488/CMP.21.43001}

\title[Low-high density state transitions in plasma-condensate systems]{Transitions from low-density state towards high-density state in stochastic bistable plasma-condensate systems}

\author[A.V. Dvornichenko, V.O. Kharchenko, D.O. Kharchenko]{A.V. Dvornichenko\refaddr{label1}, V.O. Kharchenko\refaddr{label1,label2}, D.O. Kharchenko\refaddr{label2}}

\addresses{
\addr{label1} Sumy State University, 2 Rymskyi-Korsakov St., 40007 Sumy, Ukraine
\addr{label2} Institute of Applied Physics of the National Academy of Sciences of Ukraine,\\ 58 Petropavlivska St., 40000 Sumy, Ukraine}

\date{Received August 13, 2018, in final form October 29, 2018}

\DeclareMathOperator{\mpt}{mpt}

\begin{document}

\maketitle

\begin{abstract}

In this article we study transitions from low-density states towards high-density states 
in bistable plasma-conden\-sate systems. We take into account an anisotropy in transference 
of adatoms between neighbour layers induced by the electric field near substrate. We derive 
the generalized one-layer model by assuming that the strength of the electric field is subjected to 
both periodic oscillations and multiplicative fluctuations. By studying the homogeneous system 
we discuss the corresponding mean passage time. In the limit of weak fluctuations, we show the 
optimization of the mean passage time with variation in the frequency of periodic driving in the 
non-adiabatic limit. Noise induced effects corresponding to asynchronization and acceleration in 
the transition dynamics are studied in detail.

\keywords stochastic systems, bistable systems, mean passage time

\pacs 05.10.Gg, 05.40.-a, 05.45.-a

\end{abstract}

\section{Introduction}

Plasma-condensate systems serve a useful technique to produce well structured
thin films with separated multi-layer adsorbate islands of nano-meter size of semiconductors 
and metals \cite{Perekrestov,Perekrestov2}. 
Nowadays, such nano-structured thin films attract an increased interest due to their technological 
applications in modern nano-electronic devices possessing exceptional functionality \cite{b1,b2,b3,b4}. 
Adsorptive bistable systems manifest a stochastic resonance phenomenon 
under conditions of periodically varying pressure of gaseous atmosphere (see, for example, \cite{CEJP2012}), 
or chemical potential \cite{EPJB2016}. In the technological applications, this stochastic resonance 
effect is used to optimize the output signal-to-noise ratio, when fluctuations (noise) play a constructive 
role and enhance a response of a nonlinear dynamical system subjected to a weak 
external periodic signal \cite{SR1,SR2,SR3,SR4,SR5,SR6,SR7}. 

Previously, it was shown that one can control 
the adsorbate concentration on the substrate and the corresponding first-order phase transitions 
between low-density and high-density states by varying temperature, adsorption and desorption rates
in adsorptive systems   \cite{PhysScr2012,PRE12,SS14,ME94,HM96,BHKM97,HME98_1,HME98_2}.  
At the same time, it was shown that multi-layer systems manifest a cascade of first-order phase transitions 
when a new additional layer of adsorbate is formed \cite{SS2015,Casal2002}. Such systems were mainly 
studied under the assumption of the equiprobable transference of adatoms between neighbour layers 
according to the standard vertical diffusion mechanism \cite{Wolgraef2003,Wolgraef2004}.

A fabrication of nanostructured thin films with the help of plasma-condensate devices is governed 
by the following mechanism. Ions, sputtered by magnetron, attain a growing surface 
located in a hollow cathode due to the presence of the electric field near it and become adatoms. 
Being exposed under the action of an electric field near the substrate, the main part of adatoms 
are re-evaporated to be later ionized again and to return back onto the upper layers of the 
growing surface \cite{Per8}. Hence, plasma-condensate systems are characterized by the 
anisotropy in the transitions of adatoms between neighbour layers induced by the electric field, 
with preferential motion from bottom layers towards top layers. We have shown earlier that in such 
a bistable system, the anisotropy strength related to the strength of the static electric field near 
the substrate controls the dynamics and morphology of the growing adsorbate structures \cite{NRL2017}. 
In \cite{EPJB2018}, we have derived a reduced one-layer model describing the pattern formation 
on the intermediate layer of a multi-layer system.

In the present study, we  focus on the transitions from low-density states towards high-density states 
in the effective reduced model of plasma-condensate system derived in \cite{EPJB2018}, 
by taking into account both periodic oscillations and fluctuations of the strength of the electric field. 
The main aim of the work is to define  such an external impact onto the mean passage 
time from low- to high-density state in a homogeneous model of multi-layer plasma-condensate systems.

We organize our work in the following manner. In the next section we discuss the stochastic 
model of a plasma-condensate system. In section~\ref{sec3} we analyze an influence of the periodic driving 
and stochastic force onto the mean passage time. Main conclusions 
are collected in the last section.

\section{Model of adsorptive system}

By considering the adsorbate concentration $x_n\in[0,1]$ on the selected 
$n$ layer of a multi-layer system (where $x_0=1$ corresponds to the substrate) 
we  follow \cite{SS2015,NRL2017} and describe an evolution 
of the adsorbate on each $n$-th layer by the reaction force $f(x_n)$ which includes 
adsorption, desorption and transference reactions between neighbour layers. 
Adsorption processes on any $n$-th layer are governed by the term 
$f_\text{a}=k_\text{a} p x_{n-1}(1-x_n)(1-x_{n+1})$, where 
the adsorption rate $k_\text{a}$ is defined through the adsorption energy $E_\text{a}$, temperature $T$ 
measured in energetic units and frequency factor $\nu$ as $k_\text{a}=\nu \re^{-E_\text{a}/T} $;
$p$ is the density of the plasma.
Adsorption is possible on free sites on the current $n$-th layer if  
both non-zero adsorbate concentration on the precursor ($n-1$)-th layer and free space 
on the next ($n+1$)-th layer exist. Desorption processes are described by the term 
$f_\text{d}=-k_{\text{d}n}x_n x_{n-1}(1-x_{n+1})$,
where the desorption rate $k_{\text{d}n}=k_\text{d}^0\exp[U_n(r)/T]$ is defined through the desorption
rate for non-interacting particles $k_\text{d}^0$, and interaction potential of adsorbed particles 
$U_n(r)$ giving contribution due to a strong local bond (substratum-mediated interactions). 
Desorption rate $k_\text{d}^0=\nu \re^{-E_\text{d}/T}$ relates to the life time scale of adatoms 
$\tau_\text{d}= (k_\text{d}^0 )^{-1}$, where $E_\text{d}$ is the desorption energy.
Desorption processes on $n$ layer require a non-zero adsorbate concentration on both $n$-th 
and $(n-1)$-th layers and free space on the $(n+1)$-th layer.
Transference of adatoms between neighbour layers is described by the 
ordinary vertical diffusion $f_\text{v}=D_0\left(x_{n-1}+x_{n+1}-2x_n\right)$ 
with diffusion coefficient $D_0$.
As far as in plasma-condensate devices, the electrical field presence near the substrate 
leads to the process of desorption --- additional ionization --- adsorption onto upper layers, 
we take into account such an electrical field induced motion from lower to upper layers 
in the form of additional transference of adsorbed particles from lower towards upper layers:
$k_\text{r}\left[(1-x_n)x_{n-1}-x_n(1-x_{n+1})\right]$, where the rate constant $k_\text{r}$ defines the 
anisotropy strength proportional to the strength of the electric field 
near the substrate. Formally, the electrical field should go into the exponent within the discrete state model.
In our consideration, we use a weak field approximation (in the lowest order expansion).

For the functional form of $U_n(x)$, we assume an attractive (as indicated earlier,
substratum-mediated) potential among particles separated by a distance $r$. 
In the framework of self-consistent approximation, the potential $U_n(r)$ on any 
$n$-th layer can be represented as \cite{SS2015,Casal2002,NRL2017,EPJB2018}:
\begin{equation}
U_n(r)=x_{n-1}\left[-\int u(r-r') x_n(r') \rm{d}r'\right],
\end{equation}
where the integration is provided over the whole surface. 
For the attractive potential $u(r-r')$, we assume that it is the same in any $n$-th layer. 
Following \cite{HME98_1,HME98_2} for $u(r)$, we choose a Gaussian profile
\begin{equation}
u(r)=\frac{2\epsilon}{\sqrt{4\piup r_0^2}}\exp\left(-\frac{r}{4r_0^2}\right),
\end{equation} 
where $\epsilon$ is the interaction strength and 
$r_0$ is the interaction radius. If the interaction radius is small compared
to the diffusion length, and the coverage is not much affected by variations in this
radius, we can use an approximation
\begin{equation}
\int u(r-r')x_n(r') {\rm d}r'\simeq \int u(r-r') \times \sum_n\frac{(r-r')^n}{n!}\nabla^nx_n(r){\rm d}r',
\end{equation}
which in the homogeneous case gives $U(r)\simeq-2\epsilon x_n x_{n-1}$.

To define the adsorbate concentration on both ($n-1$)-th and ($n+1$)-th layers 
through $x_n$, we  exploit the recipe proposed in \cite{EPJB2018} where the adsorbate concentration on the 
$n$-th layer, $x_n$, can be defined as the ratio between square occupied by the adsorbate on the $n$-th layer and 
on the substrate as $x_n\simeq S_n/S_0$. In accordance with the principle of surface energy minimization,  
the adsorbate concentration on each next layer of a multi-layer system is less than one on
the previous layer. By considering a multi-layer adsorbate island as a pyramidal structure with the terrace 
width $d$, the linear size of the multi-layer adsorbate structure $R_n\propto\sqrt{S_n}$ on each $n$-th layer 
decreases with the layer number $n$ growth, $R_n = R_0-nd$.
Hence, for the adsorbate concentration on each $n$ layer, we get $x_n=[1-n(d/R_0)]^2$.
By defining $x_{n-1}$ and $x_{n+1}$ in 
the same manner and introducing a small parameter $\beta_0=2d/R_0$ 
we get $x_{n\mp1}=\left(\sqrt{x_n}\pm1/2\beta_0\right)^2$. 
Next, we  measure time in units $k_\text{d}$, introduce dimensionless quantities 
$\alpha\equiv k_\text{a}/k_\text{d}$, $u\equiv k_\text{r}/k_\text{d}$, $D\equiv D_0/k_\text{d}$, $\varepsilon=\epsilon/T$ and 
fix $\beta_0=0.1$, $D=1$. By combining 
all the terms and dropping the index $n$, we finally get the evolution equation of adsorbate 
concentration on the selected level of a multi-layer plasma-condensate system  
in the following form \cite{EPJB2018}:
\begin{equation}
{\rm d}_tx=\alpha(1-x)\nu(x)-x\nu(x)\re^{\lambda(x)}+u\gamma(x)+\frac{\beta_0^2}{2}\,,
\label{eq_det}
\end{equation}
where the following notations 
$\nu(x)=(\sqrt{x}+1/2\beta_0)^2[1-(\sqrt{x}-1/2\beta_0)^2]$,
$\lambda(x)=-2 \varepsilon x (\sqrt{x}+\beta_0/2)^2$, 
$\gamma(x)=\beta_0[(1-2x)\sqrt{x}+\beta_0/4]$ are used.
The third term in the derived one-layer model (\ref{eq_det}) corresponds to the electrical field influence 
onto the system dynamics. 

\begin{figure}[!b]
\centering
\includegraphics[width=0.45\textwidth]{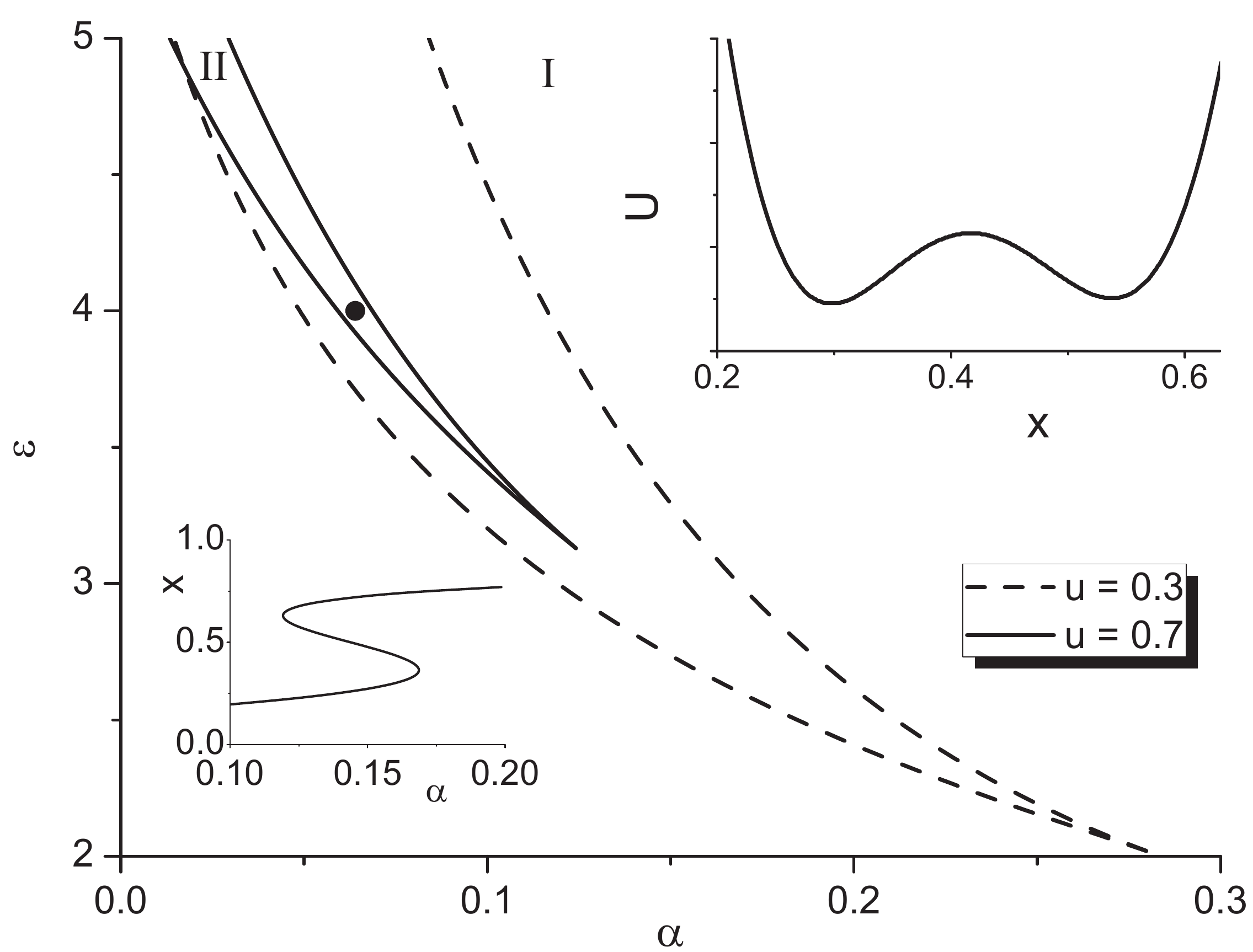}
\caption{Phase diagram for the plasma-condensate system: (I) indicates the monostability domain, 
(II) is the bistability domain.}
\label{fig2}
\end{figure}

Analysis of the stationary states of the deterministic system (\ref{eq_det}), defined from the condition 
${\rm d}_tx=0$, allows us to obtain 
the phase diagram shown in figure~\ref{fig2} that illustrates the influence of the anisotropy strength $u$ onto 
the system states. Here, in the cusp (domain II), the system is 
bistable. The bifurcation diagram $x(\alpha)$, representing first-order low-high density states 
transitions, is shown in the bottom inset at $u=0.3$, $\varepsilon=4.0$. 
It follows that an increase in the anisotropy strength shrinks the bistability domain 
in adsorption coefficient $\alpha$ and requires elevated values of the interaction strength 
$\varepsilon$ for its realization. 
The potential $U(x)=-\int f(x) {\rm d}x$ shown in the top inset corresponds to 
the spinodal and is plotted at $\alpha=0.064$, $\varepsilon=4.0$ and $u=0.7$ (black dot inside the cusp
in figure~\ref{fig2}). 

The control parameter $u$ defines the external conditions for the growth of the layers. Generally, 
it can be a function of time and/or can be changed in a stochastic manner.  
Next, we assume that the anisotropy strength~$u$ is subjected to both periodic oscillations and fluctuations: 
$u=u_0+A\sin(\omega t)+\xi(t)$, where $u_0=\langle u \rangle$ and $\xi(t)$ is the Gaussian noise 
with zero mean, $\langle\xi(t)\rangle=0$ and correlation 
$\langle\xi(t)\xi(t')\rangle=2\sigma^2\delta(|t-t'|)$; $\sigma^2$ 
is the intensity of fluctuations.
In such a case, the deterministic evolution 
equation (\ref{eq_det}) attains the form of the Langevin equation of the form
\begin{equation}
{\rm d}_tx=f(x)+A\gamma(x)\sin(\omega t)+\gamma(x)\xi(t),
\label{eq_stoch1}
\end{equation}
where $f(x)$ corresponds to the right-hand side of equation~(\ref{eq_det}). It follows that if $x=x_0$, where $\gamma(x_0)=0$, then an 
influence of the electric field near the substrate onto adsorbate concentration disappears leading to the 
isotropic deterministic system described only by adsorption, desorption and ordinary vertical diffusion. 
At the same time, $x_0$ is not an absorbing state as far as $f(x_0)\ne0$.

\section{Mean passage time} \label{sec3}

Usually, the combined effect of periodic and stochastic driving forces in bistable potentials leads to a stochastic 
resonance phenomenon. According to this scenario, a slow periodic force moves the ``Brownian particle'', 
located in one minimum of the bistable potential, towards its maximum. If the periodic driving synchronizes 
with the fluctuation force, then the last one throws the ``Brownian particle'' over the potential barrier and there takes place a switch between the 
two stable states.  Let us provide a detailed description of transitions from low- to high-density 
state by studying the passage time. To this end, we fix $\alpha=0.064$, $\varepsilon=4.0$ and $u_0=0.7$, 
corresponding to the black point on the phase diagram in figure~\ref{fig2} on the spinodal and 
perform numerical simulations of the Langevin equation 
(\ref{eq_stoch1})  with the time step $\Delta t=0.001$ 
on the graphical processor units (GPUs) with double precision. This technique provides
an effective acceleration of numerics by a factor of about 500 over the standard CPUs computing for this 
special problem. 

\begin{figure}[!b]
\centering
\includegraphics[width=0.45\textwidth]{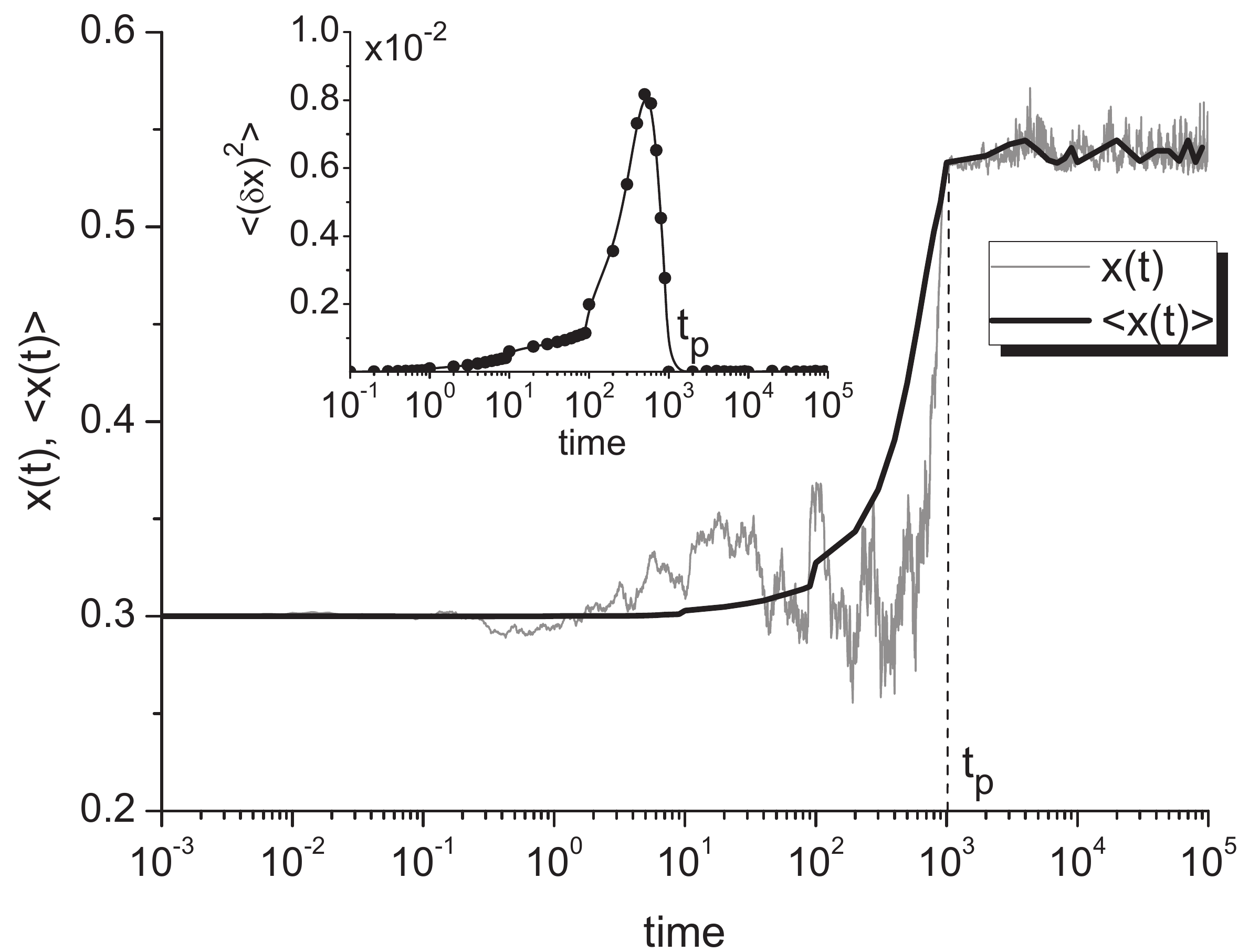}
\caption{Time dependencies of the concentration of the adsorbate (one realization shown by 
grey colour) and mean adsorbate concentration, averaged by $10^4$ realizations (black curve). 
Inset represents time dependence of the dispersion. Results were obtained at $A=0.06$ 
and $\omega=10^{-3}$, $\sigma^2=0.05$.}
\label{fig4}
\end{figure}

In figure~\ref{fig4}, we 
present the time dependence of the concentration of the adsorbate (one realization shown by 
grey colour) and mean adsorbate concentration, averaged by $10^4$ realizations (black curve).
In all simulations, the initial condition for the adsorbate concentration was selected in the minimum 
of the potential $U(x)$ corresponding to the low-density state. It follows that during the system evolution, the 
combined influence of periodic and stochastic driving leads to the transition towards 
a high-density state that in average occurs at time instant $t_\text{p}$, 
when the dispersion $\langle(\delta x)^2\rangle$, averaged over an ensemble, falls to zero after its
maximum (see inset in figure~\ref{fig4}).   
Next, we  study the influence of the periodic driving (amplitude $A$ and frequency $\omega$) 
and the stochastic force (intensity of fluctuations  $\sigma^2$) onto the mean passage time ($\mpt$) from 
the low-density state, corresponding to minimum of the bistable potential $U(x)$ with small $x$, towards the 
high-density state [the minimum of $U(x)$ with large $x$], defined as: 
$\mpt=N^{-1}\sum_{i=1}^{N}t_\text{p}$, where sum is taken over $N=10^{4}$ realizations.

\subsection{Limit of quasi-deterministic driving}

\begin{figure}[!h]
\centering
a)\includegraphics[width=0.47\textwidth]{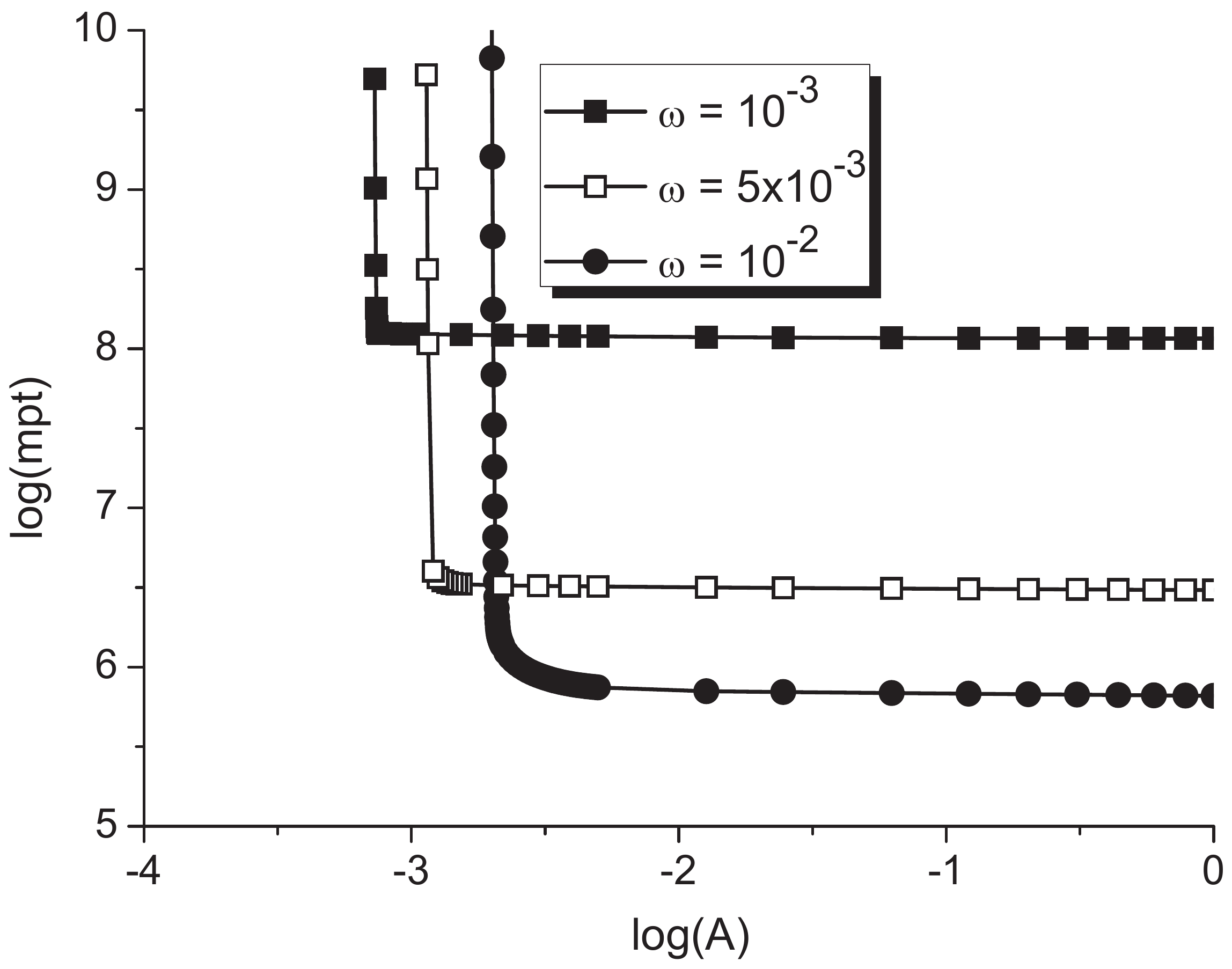}
b)\includegraphics[width=0.47\textwidth]{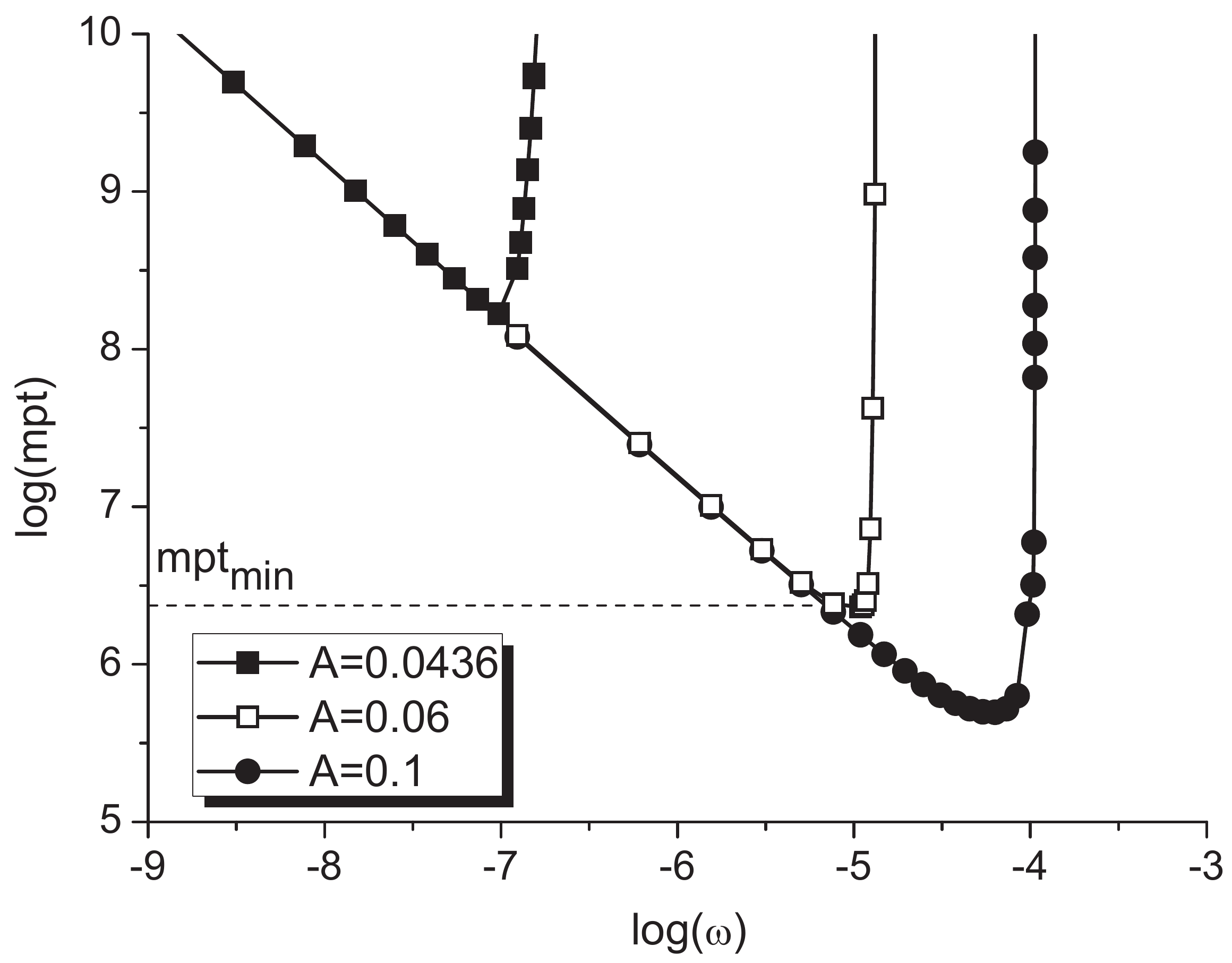}
\caption{Dependencies of the mean passage time from low-density state to high-density state on 
a) amplitude of the periodic driving $A$ at different values on the frequency $\omega$ and 
b) frequency $\omega$ at different values of amplitude of the periodic driving $A$ at 
$\sigma^2=10^{-5}$.}
\label{fig5}
\end{figure}

First, we focus our attention onto the influence of the periodic driving in the limit of weak
fluctuations with $\sigma^2=10^{-5}$. Dependencies of the mean passage time $\mpt$ 
from the low-density state to the high-density state on the amplitude of the periodic driving $A$ at 
different values of the frequency $\omega$ and on the frequency $\omega$ at different values of 
the amplitude of the periodic driving $A$ are shown in figures~\ref{fig5}~(a), (b), respectively.
From figure~\ref{fig5}~(a) it follows that at small values of the amplitude $A$, the transition becomes 
impossible due to $\log(\mpt)\to\infty$ independent of the frequency $\omega$. With an increase  in
the amplitude~$A$, the value of the $\mpt$ abruptly decreases and remains constant at large values 
of $A$. An increase in the periodic driving frequency $\omega$ requires elevated values of the 
driving amplitude $A$ for the transition, on the one hand, and results in a decrease in the transition 
time at large $A$, on the other hand. 

The dependence $\mpt(\omega)$, shown in figure~\ref{fig5}~(b) is of a more complicated structure. Here, 
we have a special kind of synchronization: at a fixed value of the periodic force amplitude $A$, an increase in 
the frequency leads to a decrease in the transition time, until the minimal value $\mpt_\text{min}$ is reached; 
at a further growth in $\omega$, the $\mpt$ increases. This minimal value of the transition time 
$\mpt_\text{min}$ decreases with the growth of the amplitude $A$. Hence, the $\mpt$ optimizes with the frequency 
of the periodic driving at non-adiabatic limit.    

\subsection{Noise-induced effects}

\begin{figure}[!t]
\centering
a)\includegraphics[width=0.47\textwidth]{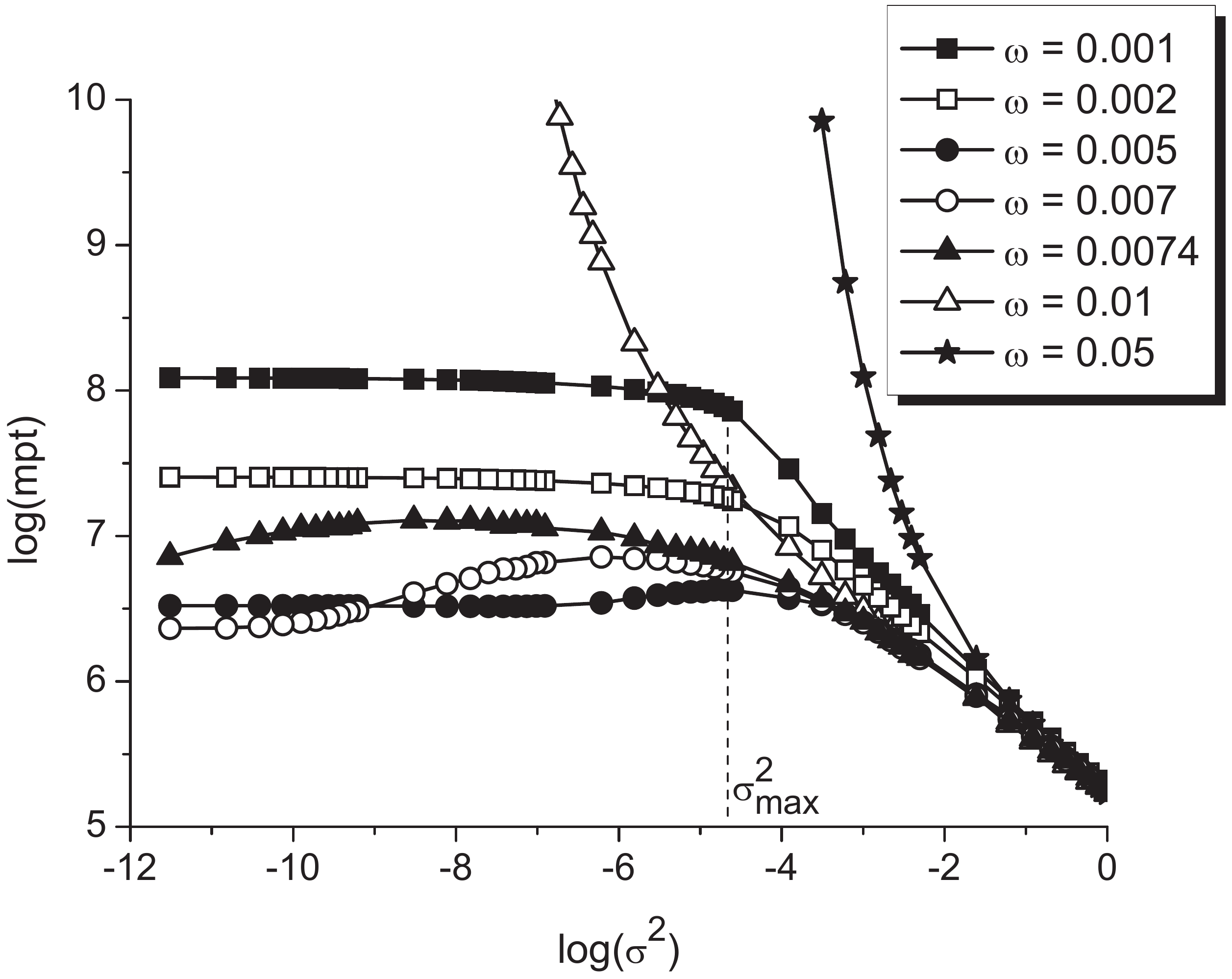}
b)\includegraphics[width=0.47\textwidth]{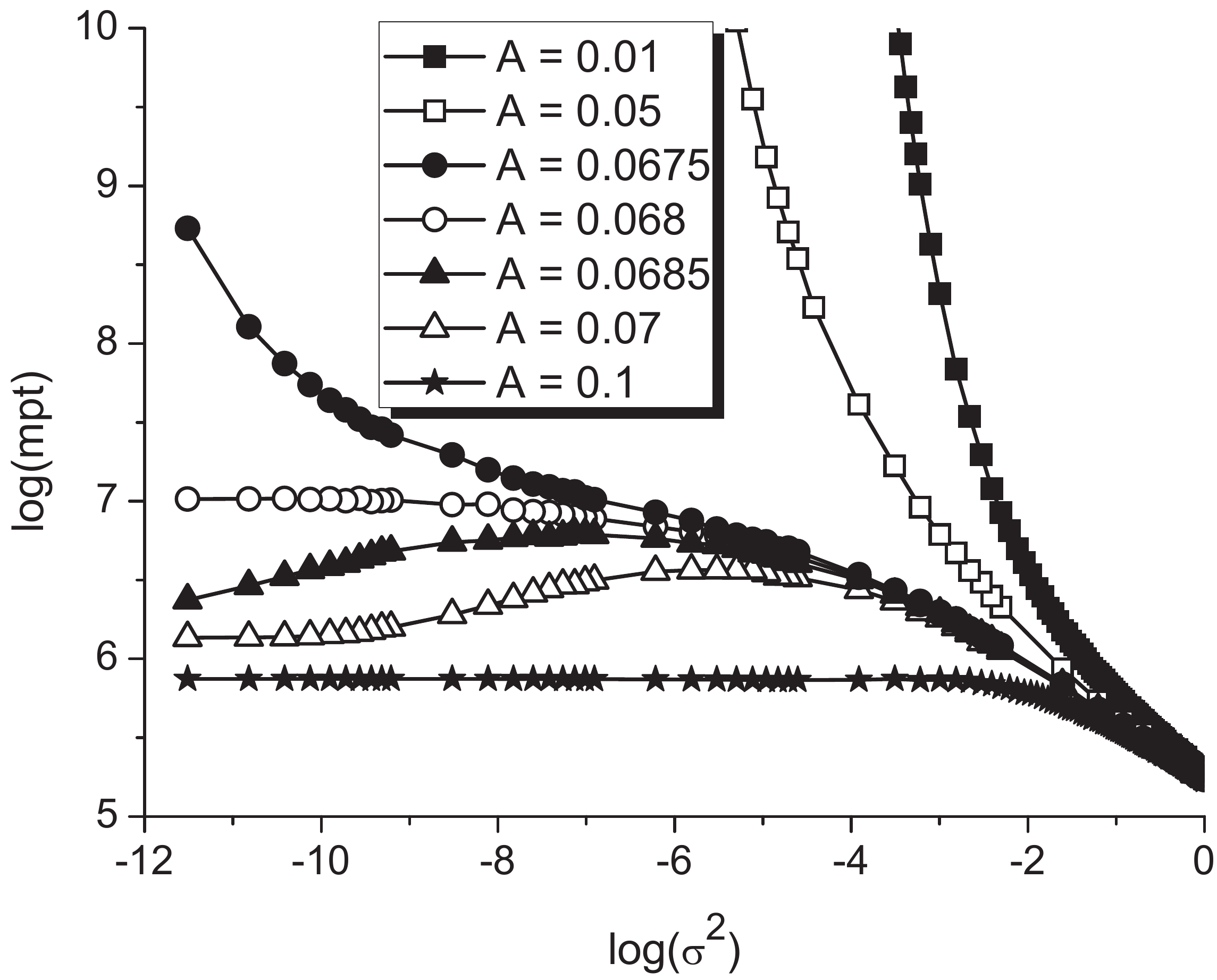}
\caption{Dependencies of the mean passage time from low-density state to high-density state on 
 intensity of fluctuations $\sigma^2$ at: 
a) $A=0.06$ and different values of the frequency $\omega$ of periodic driving;
b) $\omega=0.01$ and different values of the amplitude of the periodic driving $A$.}
\label{fig6}
\end{figure}

Next, we  analyze a change in the transition time $\mpt$ by varying the  intensity of fluctuations
$\sigma^2$ for different values of amplitude $A$ and frequency $\omega$ of periodic driving. 
The dependencies $\mpt(\sigma^2)$ at $A=0.06$ and different values of 
$\omega$ are shown in figure~\ref{fig6}~(a).  
It is seen that at small values of the driving frequency $\omega$, an increase in the noise intensity 
weakly decreases the transition time [see curve with filled squares at $\omega=0.001$ in figure~\ref{fig6}~(a)]. 
If $\sigma^2$ becomes large enough, 
$\sigma^2>\sigma^2_\text{max}$, then the stochastic force starts to play a dominant role 
in the system dynamics and a further increase in its intensity significantly decreases the 
transition time from low- to high-density state. An 
increase in the periodic driving frequency at small $\sigma^2$ acts in the manner presented in figure~\ref{fig5}~(b): 
$\mpt$ decreases, attains the value $\mpt_\text{min}$ and then increases. At 
$\sigma^2<\sigma^2_\text{max}$, the value $\mpt_\text{min}$ weakly increases with $\sigma^2$.  
At large values of the periodic force frequency, the transition from low- to high-density state 
occurs only at elevated values of the intensity of fluctuations. At narrow interval of frequency 
values, the mean transition time manifests a non-monotonous dependence on the noise intensity 
(see curves with filled and empty circles at $\omega=0.005$ and $0.007$ and curve with 
filled triangles at $\omega=0.0074$). Here, with an increase in the intensity of fluctuations,
the transition time increases, attains maximal value and then decreases. Hence,
at small values of the noise intensity, its increase leads to the delay in the transition dynamics. 
It means that in such conditions, one gets asynchronization in periodic and stochastic driving: 
while periodic force moves the ``Brownian particle'' towards the maximum of the bistable potential, 
fluctuations return it back to the low-density state.
With a further increase in $\sigma^2$, the noise starts to play the dominant role in the system 
dynamics which leads to the transition through the potential barrier.   

In figure~\ref{fig6}~(b), we present dependencies of the mean transition time on the noise intensity 
at a fixed value of the periodic driving frequency $\omega=0.01$ and different values of the driving 
amplitude $A$. It follows that with an increase in $A$, the transition from low- to high-density state 
occurs faster. An increase in the noise intensity leads to (i) a decrease in the transition time at small 
values of the periodic driving amplitude; (ii) a delay in the transition dynamics at intermediate values 
of $A$; (iii) at elevated values of the periodic driving amplitude, the noise influences the system dynamics only 
at large values.  

\section{Conclusions}

In this article, we have provided a detailed study of the transitions from low-density state 
to high-density state in multi-layer plasma-condensate systems in a reduced one-layer model. 
By taking into account the anisotropy in transference between layers, induced by the electric field 
near the substrate, we assume periodic oscillations and fluctuations of the electric field strength. 
We discussed the influence of both periodic driving and stochastic force onto the mean passage 
time needed for transition from the low-density state towards the high-density state.
It is shown that in the case of a weak fluctuating force, the mean passage time decreases with the 
periodic driving amplitude growth and optimizes with the frequency of periodic driving.
An increase in the intensity of the electrical field fluctuations at small $\sigma^2$ delays 
the transition towards the high-density state due to asynchronization in periodic and stochastic driving; 
a growth in the intensity of fluctuations at large $\sigma^2$, leads to a decrease in the time needed to 
pass from low-density state towards high-density state. 

We expect that this investigation could be also useful for a source of hydrogen
negative ions with a metal-hydride electrode \cite{add1} and while using the metal-hydride
as a plasma-facing material in fusion reactors~\cite{add2}.

\section*{Acknowledgements}

Support of this research by the Ministry of Education and Science of Ukraine, 
project \linebreak No.~0117U003927, is gratefully acknowledged.

\newpage

\ukrainianpart

\title{Переходи від стану з низькою концентрацією до стану з високою концентрацією у 
стохастичній бістабільній системі плазма-конденсат}%
\author{А.В. Дворниченко\refaddr{label1}, В.О. Харченко\refaddr{label1,label2}, Д.О. Харченко\refaddr{label2}}
\addresses{
\addr{label1} Сумський державний університет, вул. Римського-Корсакова, 2, 40007 Суми, Україна 
\addr{label2} Інститут прикладної фізики НАН України, вул. Петропавлівська, 58, 40000 Суми, Україна}

\makeukrtitle

\begin{abstract}
\tolerance=3000%

У цій статті нами досліджено переходи від стану з низькою густиною адсорбату до стану 
з  його високою концентрацією в бістабільних системах плазма-конденсат.  При цьому враховується 
анізотропія у переходах адатомів між сусідніми шарами, що спричинена дією 
електричного поля поблизу поверхні. В рамках узагальненої одношарової моделі досліджуються ефекти, 
пов'язані з періодичними коливаннями та флуктуаціями напруженості електричного поля.   
При дослідженні однорідної системи встановлено особливості зовнішнього впливу на середній 
час переходу від стану з низькою густиною адсорбату до стану 
з високою його концентрацією. В границі слабких флуктуацій продемонстровано процес оптимізацій середнього часу переходу 
при зміні частоти періодичного керуючого поля. Детально вивчаються ефекти, пов'язані з процесами асинхронізації та прискорення в динаміці переходів.

\keywords
 стохастичні системи, бістабільні системи, середній час переходу між станами
\end{abstract}

\end{document}